\documentclass{iaus}
\usepackage{graphicx}

\title[High-res view of planet formation sites] 
{High-resolution spectroscopic view of planet formation sites}

\author[Zs. Reg\'aly \& L. Kiss \& Zs. S\'andor \& C. P. Dullemond]   
{Zs. Reg\'aly$^1$,
L. Kiss$^{1,2}$,
Zs. S\'andor$^3$
\and C. P. Dullemond$^3$
}

\affiliation{$^1$Konkoly Observatory of the Hungarian Academy of Sciences, P.O. Box 67, H-1525 Budapest, Hungary \\ email: {\tt regaly@konkoly.hu} \\[\affilskip]
$^2$Sydney Institute for Astronomy, School of Physics, University of Sydney, Australia\\[\affilskip]
$^3$Junior Research Group, Max-Planck-Institut f\"ur Astronomie, K\"onigstuhl 17, D-69117 Heidelberg, Germany
}

\pubyear{2010}
\volume{276}  
\pagerange{119--126}
\jname{The Astrophysics of Planetary Systems: Formation, Structure, and Dynamical Evolution}
\editors{A. Sozzetti, M. G. Lattanzi \& A. P. Boss}
\begin{document}

\maketitle

\begin{abstract}
Theories of planet formation predict the birth of giant planets in the inner, dense, and gas-rich regions of the circumstellar disks around young stars. These are the regions from which strong CO emission is expected. Observations have so far been unable to confirm the presence of planets caught in formation. We have developed a novel method to detect a giant planet still embedded in a circumstellar disk by the distortions of the CO molecular line profiles emerging from the protoplanetary disk's surface. The method is based on the fact that a giant planet significantly perturbs the gas velocity flow in addition to distorting the disk surface density. We have calculated the emerging molecular line profiles by combining hydrodynamical models with semianalytic radiative transfer calculations. Our results have shown that a giant Jupiter-like planet can be detected using contemporary or future high-resolution near-IR spectrographs such as VLT/CRIRES or ELT/METIS. We have also studied the effects of binarity on disk perturbations. The most interesting results have been found for eccentric circumprimary disks in mid-separation binaries, for which the disk eccentricity - detectable from the asymmetric line profiles - arises from the gravitational effects of the companion star. Our detailed simulations shed new light on how to constrain the disk kinematical state as well as its eccentricity profile. Recent findings by independent groups have shown that core-accretion is severely affected by disk eccentricity, hence detection of an eccentric protoplanetary disk in a young binary system would further constrain planet formation theories.

\end{abstract}

\firstsection 

\section{Introduction}
\noindent
It is well known that circular Keplerian protoplanetary disks are expected to produce symmetric double-peaked molecular line profiles \cite[(Horne \& Marsh 1986)]{HorneMarsh1986}. Contrary to this simple symmetric disk assumption, asymmetric CO line profiles in the fundamental band have been observed in several cases. Grid-based numerical simulations of \cite[Kley \& Dirksen (2006)]{KleyDirken2006} have shown that local disk eccentricity might form in planet bearing disk near the gap. The theory of resonant excitation mechanisms in accretion disks of \cite[Lubow (1991)]{Lubow1991} predicts that the circumstellar disks of close-separation young binaries might become fully eccentric due to the orbiting companion. \cite[Horne (1995)]{Horne1995} has shown that supersonic turbulence might cause observable line profile distortions. As the orbital velocity of gas parcels is highly supersonic in accretion disks, the disk eccentricity results in supersonic velocity distortions. Therefore distorted molecular line profiles are expected to form in giant planet bearing locally eccentric protoplanetary disks \cite[(Reg\'aly et al. 2010)]{Regalyetal2010} and fully eccentric circumstellar disks of close-separation young binaries.

\section{Spectral calculations combined with hydrodynamic simulations}
\noindent
In order to calculate the CO spectra, we need to know the temperature distribution in the disk. In our study the double layer flaring disk model of \cite[Chiang \& Goldreich (1997)]{ChiangGoldreich1997} is assumed. In this model the disk is heated by the stellar irradiation and accretion processes. The incident stellar flux heats the disk atmosphere, which reprocesses the stellar light and irradiates the disk interior. The accretion processes heats the disk interior directly. In this way an optically thick interior and an optically thin atmosphere develop (Fig. 1). For low accretion rate, temperature inversion forms, resulting in emission spectra.

\begin{figure}[h]
	\begin{center}
		\vspace{-0.5cm}
		\includegraphics[width=10cm]{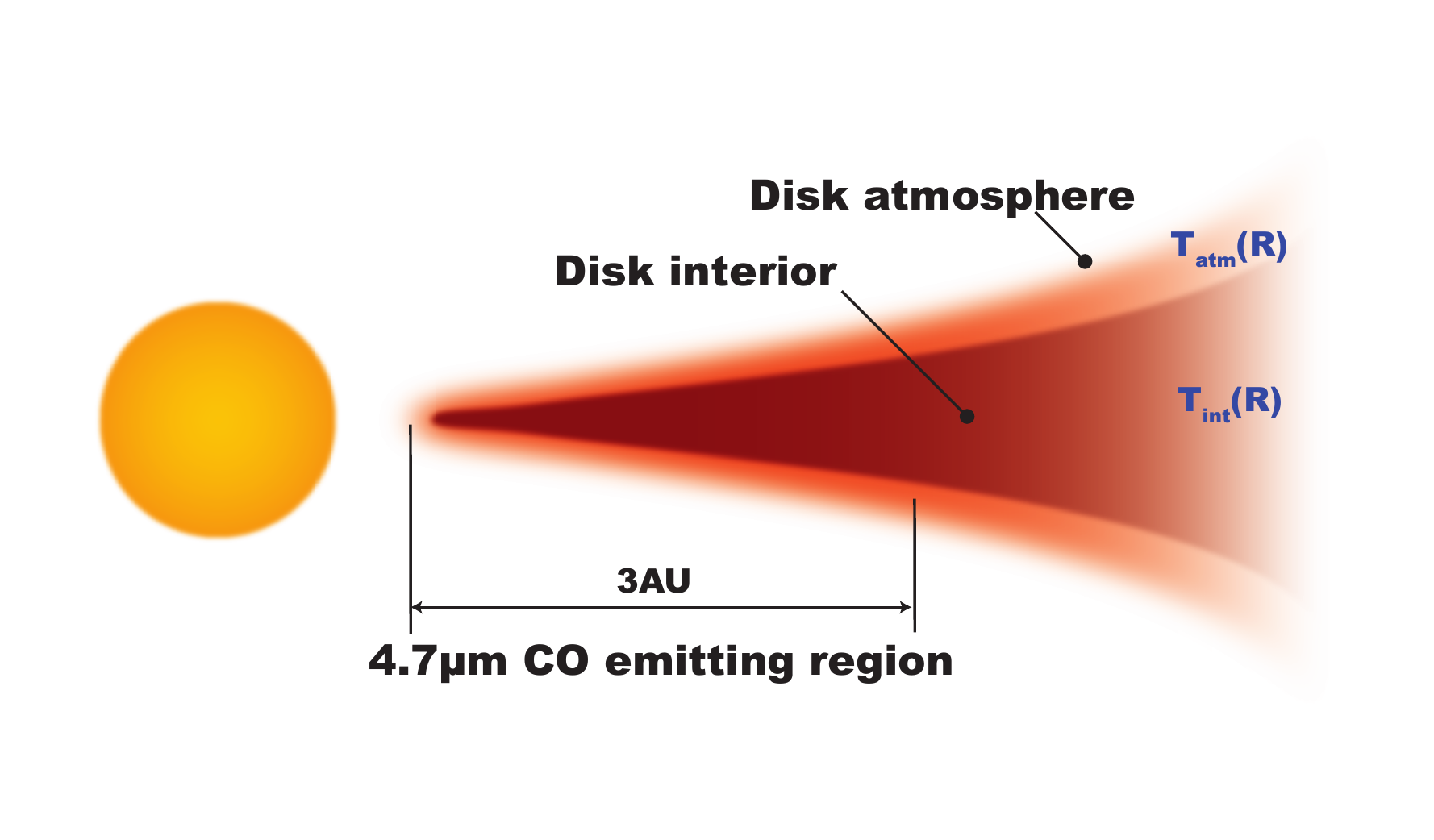} 
		\vspace{-1cm}
		\caption{The double layer disk model in which the disk atmosphere is superheated.}
		\label{figm}
	\end{center}
\end{figure}

The fundamental band CO emission is formed in the optically thin disk atmosphere above the optically thick cooler disk interior. In the optically thin approximation the $I(\nu,R,\phi)$ monochromatic intensity of the radiation at frequency $\nu$ from a given gas parcel at $R,\phi$ cylindrical coordinates is the sum of the dust continuum of the disk interior and the optically thin atmospheric CO emission
\begin{equation}
	I(\nu,R,\phi,i)=B(\nu,T_d(R))e^{-\tau(\nu,R,\phi,i)}+B(\nu,T_g(R))(1-e^{-\tau(\nu,R,\phi,i)}),
\end{equation}
where, $B(\nu,T)$ is the Planck function, $T_d(R)$ and $T_g(R)$ are the dust and gas temperatures in the disk interior and the disk atmosphere, respectively. Details of the spectral model can be found in \cite[Reg\'aly et al. (2010)]{Regalyetal2010}. For a disk perturbed by a companion (a planet or a secondary star) the velocity field is no longer circular Keplerian, in which case the local Doppler shift that affects the atmospheric optical depth ($\tau(\nu,R,\phi,i)$) can be given by 
\begin{eqnarray}
	\Delta\nu(R,\phi,i)&=&\frac{\nu_{0}}{c}\sin(i)\left\{\left[u_\mathrm{R}(R,\phi)\cos(\phi)-u_\mathrm{\phi}(R,\phi)\sin(\phi)\right]\right.  \nonumber\\
	&&+\left.\left[u_\mathrm{R}(R,\phi)\sin(\phi)+u_\mathrm{\phi}(R,\phi)\cos(\phi)\right]\right\},
	\label{eq:Doppler-shift-pertdisk}
\end{eqnarray}
where $u_\mathrm{R}(R,\phi)$ and $u_\phi(R,\phi)$ are the radial and the azimuthal velocity components of the orbiting gas parcels. In order to calculate the local Doppler shift in a perturbed disks with embedded planet or a secondary star, we solved the continuity and Navier-Stokes equations by the 2D grid-based hydrodynamic code FARGO \cite[(Masset 2000)]{Masset2000}. In this way, the $u_\mathrm{R}(R,\phi)$ and $u_\phi(R,\phi)$ velocity components of gas parcels, were provided by the hydrodynamic simulations. For simplicity we used a locally isothermal equation of state for the gas. The $\alpha$-type disk viscosity was applied \cite[(Shakura \& Sunyaev 1987)]{ShakuraSunyaev1987}. The disk self-gravitation was neglected as the Toomre  parameter is well above 1. For more details on the hydrodynamic simulations see \cite[Reg\'aly et al. (2010)]{Regalyetal2010}.

\section{Planetary signal: local disk eccentricity due to a giant planet}
\begin{figure}[h]
	\begin{center}
		\includegraphics[width=\columnwidth]{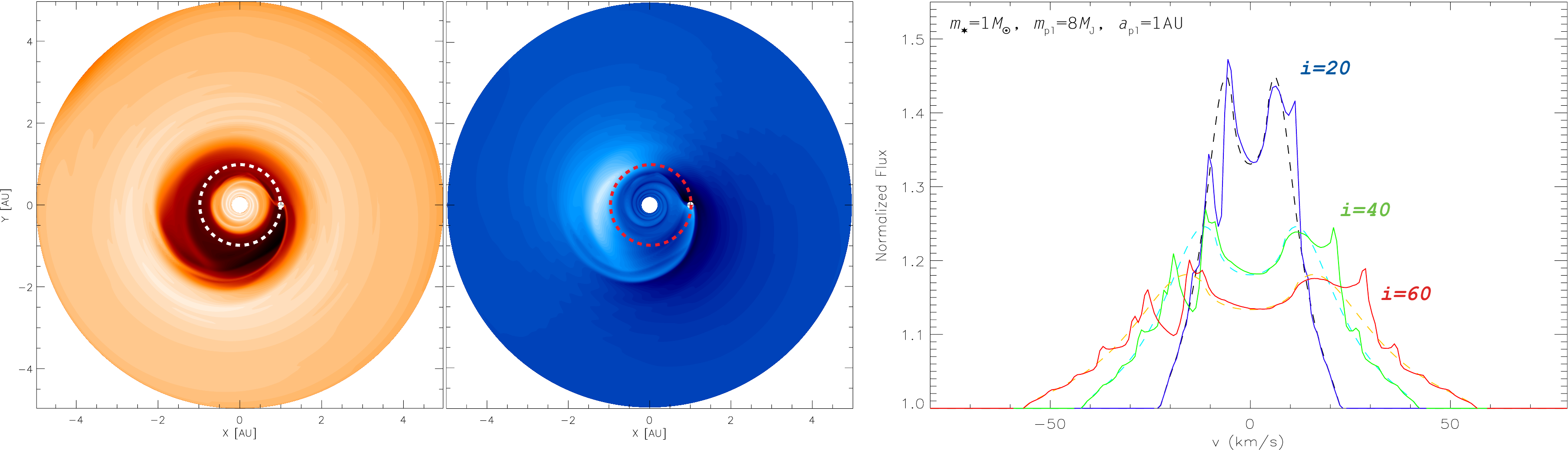} 
		\caption{Surface density (\emph{left panel}) and radial component of velocity (\emph{middle panel}) distribution in a planet bearing protoplanetary disk. Distorted $V=1\rightarrow0\,\mathrm{P10}$ CO lines (\emph{right panel}) emerging from the gravitationally distorted disk, assuming $20^\circ$, $40^\circ$ and $60^\circ$ disk inclinations.}
		\label{fig1}
	\end{center}
\end{figure}

\noindent
According to our simulations the density distribution shows permanent elliptic shape near the gap (Fig. \ref{fig1}, left panel), confirming the results of \cite[Kley \& Dirksen (2006)]{KleyDirken2006}. We found that the orbits of gas parcels are eccentric in the vicinity of the planet (Fig. \ref{fig1}, middle panel). As the magnitude of the velocity distortion of the gas parcels exceeds the local sound speed, distorted non-symmetric line profiles are expected to form.

We have calculated the fundamental band P10 non-blended CO line for 20, 40 and 60 degree of disk inclination in a perturbed disk assuming an $8M_\mathrm{Jup}$ mass planet orbiting $1\,M_\odot$ mass star at 1\,AU (Fig. \ref{fig1}, right panel). The line shape is distorted and the relative strength of the distortions are increasing with increasing inclination angle. Calculating the line profiles with different orbital positions of the planet, we have found that the line shapes vary as the planet is orbiting.

In order to investigate the dependence of the line profile distortions on the model parameters, we have calculated the CO emission in several models assuming different planetary and stellar masses, and orbital distance of the planet. We found that the planetary and the stellar mass affect the line profiles by the same means: the larger the planetary or stellar mass, the stronger the distortion. The line profile distortion is found to be weakening with increasing orbital distance. According to our calculations, the signal of a Jovian planet orbiting a Solar mass star at a distance of 1\,AU can be detected with CRIRES (Reg\'aly et al., 2010).

\section{Asymmetric lines: fully eccentric disk due to a secondary star}
\noindent
Circumprimary disks in young binaries become fully eccentric via the gravitational perturbations of the secondary, in form of resonant excitation mechanisms (\cite[Lubow 1991]{Lubow1991}, \cite[Kley et al. 2008]{Kleyetal2008}). \cite[Th\'ebault et al. (2006)]{Thebaultetal2006} and \cite[Paardekooper et al. (2008)]{Paardekooperetal2008} have shown that the planetesimal accretion might be inhibited by the disk eccentricity in core-accretion models. Recently, Zsom et al. (2010) have shown that the dust coagulation process is also affected by the disk eccentricity in the core-accretion scenario.

\begin{figure}[h]
	\begin{center}
		\includegraphics[width=\columnwidth]{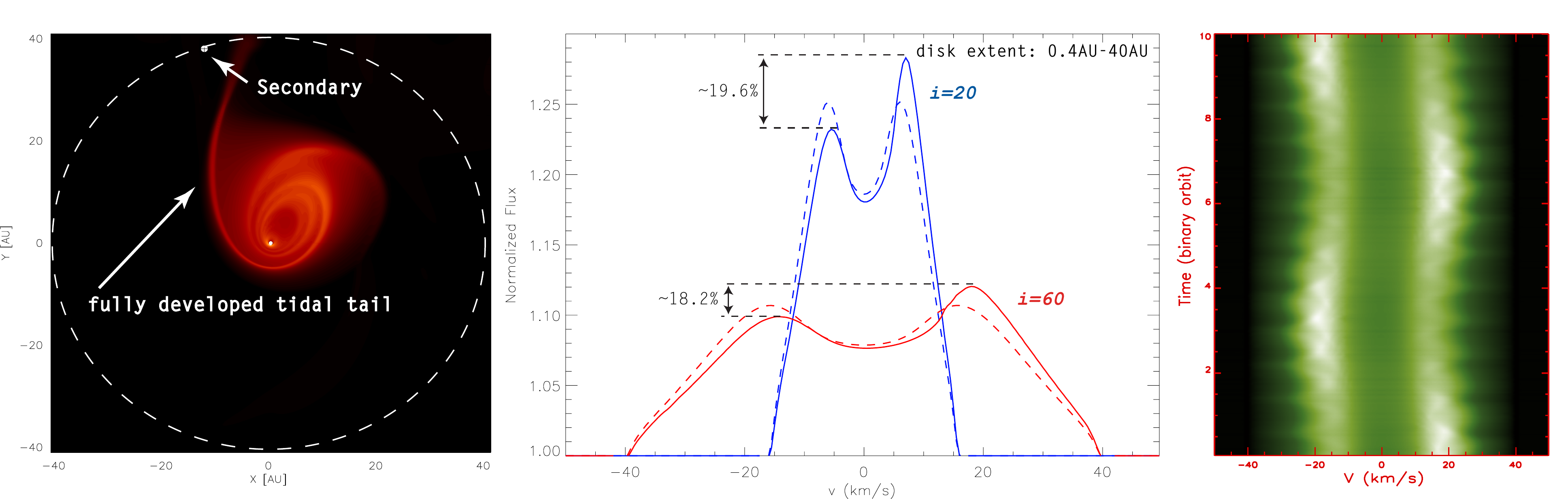} 
		\caption{The circumprimary disk becomes eccentric due to the perturbations from the secondary. Asymmetric $V=1\rightarrow0\,\mathrm{P10}$  CO  line (\emph{middle panel}), assuming $20^\circ$ and $60^\circ$ disk inclinations. Trailed spectra (\emph{right panel}) covering a full disk precession shows the periodic variations in asymmetry with $\sim 6.5 P_\mathrm{bin}$ and line wings with $\sim P_\mathrm{bin}$.}
		\label{fig2}
	\end{center}
\end{figure}

According to our simulations, the average disk eccentricity is $\sim0.3$ independently on the binary mass ratio and the magnitude of viscosity, assuming $h=0.05$ aspect ratio for the disk, which is a reasonable assumption for a circumstellar disk around an average T\,Tauri star (Fig. \ref{fig2}, left panel). However, smaller disk eccentricity was found for thick ($h\geq0.75$) and thin ($h \leq0.03$) disks. We have found that the orbital eccentricity of the binary above 0.2 inhibits the development of disk eccentricity.

Calculating the $V=1\rightarrow0\,\mathrm{P10}$ non-blended CO lines with $20^\circ$ and $60^\circ$ disk inclination angle, we have found a maximum $\sim20$\% red-blue peak asymmetry (Fig. \ref{fig2}, middle panel). The profile asymmetry is due to the fact that the inner disk, where the fundamental band CO emission is formed, is fully eccentric. The peak asymmetry is inverted twice during a full disk precession, which takes about $\sim6.5P_\mathrm{bin}$. Moreover, as the disk eccentricity slightly varies during one binary orbit, line profile variations might be observed in the line wings with $\sim P_\mathrm{bin}$.

\section{Summary}
\noindent
High-resolution 4.7$\mu$m CO molecular spectra provide us a tool to reveal disk eccentricity within the 3\,AU regions, where terrestrial planets are expected to form. We have shown that Jovian planets still embedded in their protoplanetary disk can indirectly be detected with the 4.7\,$\mu$m CO line profile distortions caused by the local disk eccentricity. We have also shown that significant CO line profile asymmetry and variations are expected from fully eccentric disks of close separation young binaries. By revealing disk eccentricity in close-separation young binaries the core-accretion model might be constrained further, since the planet formation seems to be strongly affected by the disk eccentricity.

\vskip 0.3cm
\noindent
{\small {\bf Acknowledgements:} This project has been supported by the "Lend\"ulet" program of the Hungarian Academy of Sciences and the DAAD-PPP mobility grant P-M\"OB/841/.}





\end{document}